\documentclass[10pt,onecolumn,a4paper]{article}
\newcommand{\affil}[1]{$^{\rm #1}$}
\usepackage[authoryear]{natbib}
\bibpunct{(}{)}{;}{a}{}{,}
\usepackage{graphicx}
\usepackage{psfig}
\date{} 
\title{Pulling out Threads from the Cosmic Tapestry: \\
Defining Filaments of Galaxies}
\author{{\it Kevin A. Pimbblet\affil{A,B}}\\\\
\affil{A}\,Department of Physics, University of Queensland, Brisbane, QLD 4072, Australia\\
\affil{B}\,E-mail: pimbblet@physics.uq.edu.au}
\begin{document}
 \maketitle
\begin{minipage}{.9\textwidth}
{\bf Abstract}\\
%
Filaments of galaxies are the dominant feature of modern large
scale redshift surveys.  They can account for up to perhaps half
of the baryonic mass budget of the Universe and 
their distribution and abundance can help
constrain cosmological models.  However, there remains no
single, definitive way in which to detect, describe and define 
what filaments are and their extent.  This work examines a
number of physically motivated, as well as statistical,
methods that can be used to define filaments and
examines their relative merits.

\medskip
\end{minipage}

%
%
\section{Motivation}
What is a filament of galaxies (FOG)?  
Although at first glance, this is a seemingly innocuous,
benign and near-trivial
question, there is not really a straight forward answer to it.
Many authors, including the present one, have recently been
searching for more concrete definitions and hence, also, methods 
of finding and detecting FOGs (e.g.\ Pimbblet 2005 and references
therein).

The present work is therefore a timely review of the 
(growing) myriad of approaches 
that exist to define and detect FOGs in an attempt to answer the
question of what a FOG actually is.
In a lot of ways, investigations of FOGs nowadays
are arguably analogous to where the investigations of galaxy clusters 
stood at about one half of a century ago (see Abell 1965 for 
an excellent, albeit somewhat dated by todays
standards, review of galaxy clustering).
Undoubtedly, the reason for the recent flurry of activity into
investigating and characterizing
FOGs has to be the availability of modern, high-quality
and, most importantly,
wide-field redshift surveys such as the 2dF Galaxy Redshift Survey
(2dFGRS; e.g.\ Colless et al.\ 2001), the Sloan Digital Sky
Survey (SDSS; e.g.\ Abazajian et al.\ 2004),
the 6dF Galaxy Survey (6dFGS; e.g.\ Jones et al.\ 2004) 
and the Las Campanas
Redshift Survey (LCRS; e.g.\ Shectman et al.\ 1996).

\begin{figure}
\centerline{\psfig{file=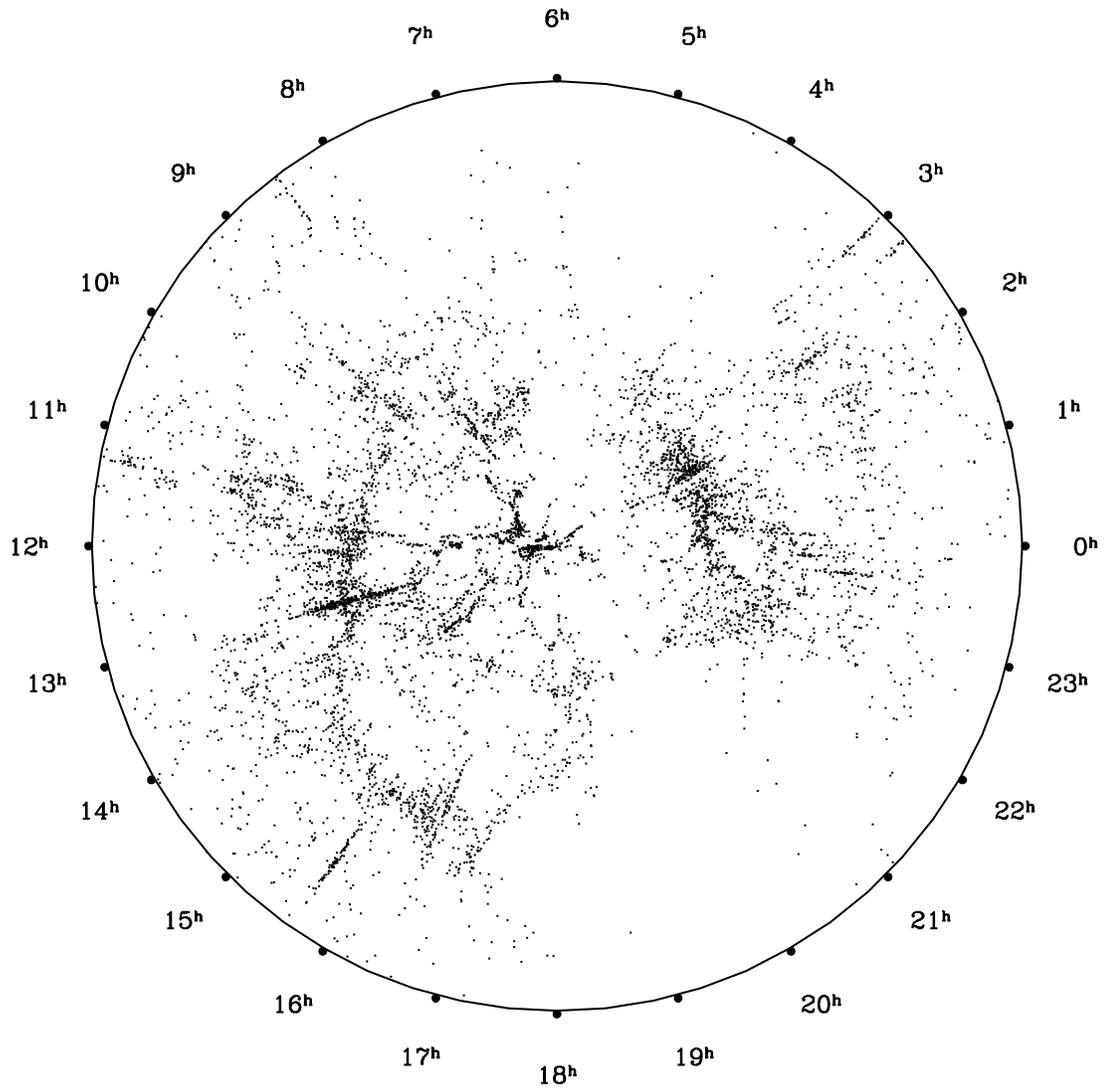,angle=0,width=6.in}}
  \caption[The Great Wall]{\small{The Great Wall of Geller \& Huchra (1989) 
reconstructed using J2000 coordinates from the November 1993 
public data release of
the CfA dataset.  Following Geller \& Huchra (1989), all galaxies within
a declination range of $20 < \delta < 40$ are plotted with no cut in 
magnitude made.  The radius of the circle is 15000 km s$^{-1}$.
The Great Wall can be seen extending outward
from 13 hours.
}}
  \label{fig:gw}
\end{figure}

Having written that, the discovery of 
significant segments of large-scale structure other than
galaxy clusters
-- sheets, filaments and walls of galaxies --
is not a new thing.  Geller \& Huchra (1989) famously
cartographed the `Great Wall' from the CfA redshift survey
(e.g.\ Huchra et al.\ 1983): a highly significant
feature that stretches for at least $170 \times 60 h^{-1}_{100}$ Mpc
at $cz \approx 7500$ km s$^{-1}$ (Figure~\ref{fig:gw}).
So significant is this detection, that it should even have an
imprint on the cosmic microwave background radiation
(Atrio-Barandela \& Kashlinsky 1992; see also Chodorowski 1994).

%

\begin{figure}
\centerline{\psfig{file=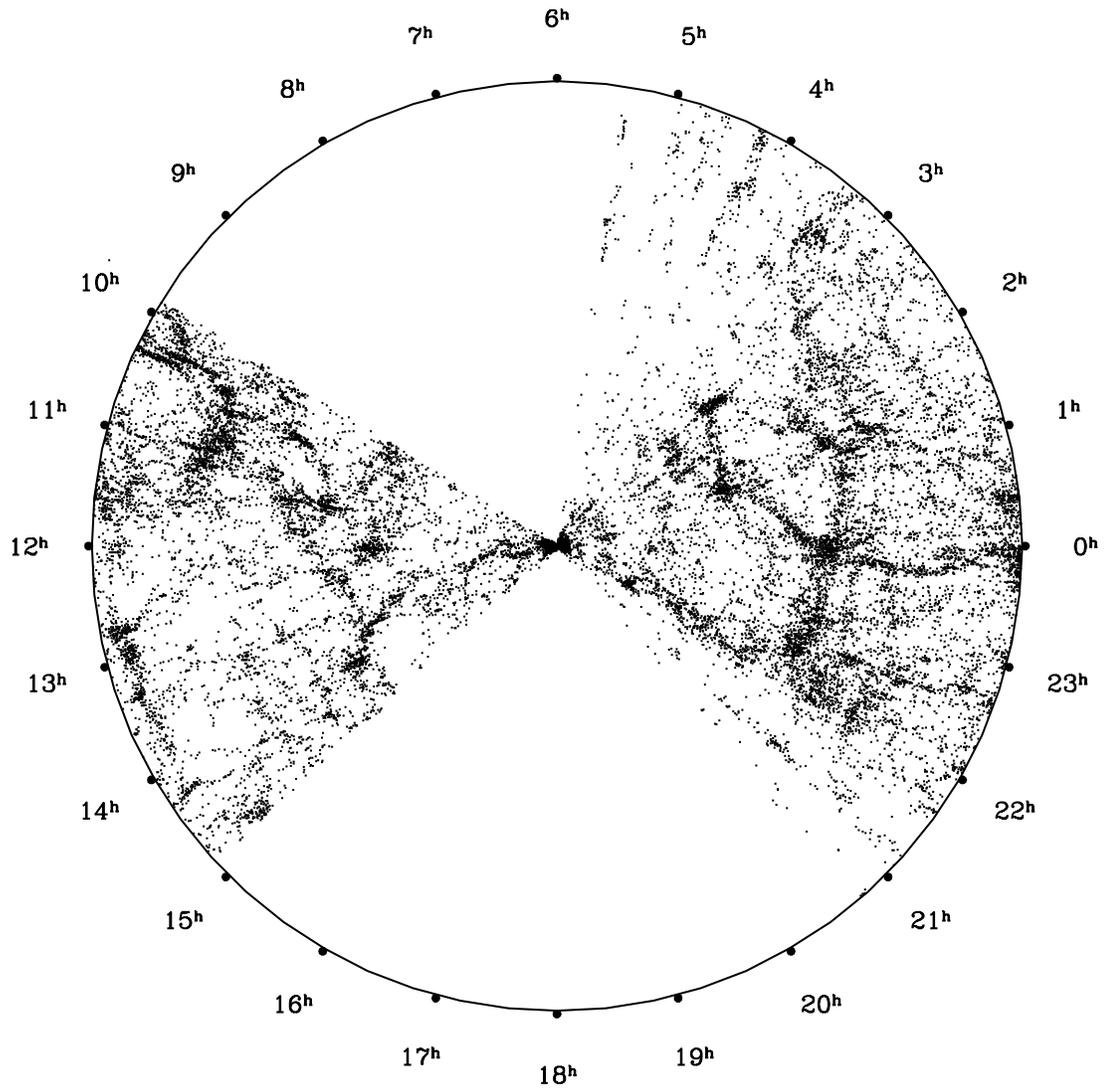,angle=0,width=6.in}}
  \caption[The Great Wall]{\small{As for Figure~\ref{fig:gw},
but using data from 2dFGRS.  No cut has been made in declination or
$b_J$ magnitude, although the plot has been constrained to 
the same redshift limit as Figure~\ref{fig:gw}, albeit with
a different declination range.  
Note the complex manner in which FOGs 
weave through the structure and its qualitative
similarity to the CfA survey.
}}
  \label{fig:twodf}
\end{figure}

What else may we expect from FOGs?  We know that 
in hierarchical structure formation modeling 
there has long been the prediction that
galaxy clusters grow through repeated mergers with other 
galaxy clusters (and galaxy groups)
together with continuous accretion of their surrounding matter 
(e.g.\ Zeldovich, Einasto \& Shandarin 1982; Katz et al.\ 1996;
Jenkins et al.\ 1998; Colberg et al.\ 2000;
see also Bond, Kofman \& Pogosyan 1996).
We also know that the accretion process occurs in
a highly non-isotropic manner: galaxy filaments funnel matter onto 
large clusters along preferred directions (see 
Pimbblet 2005; Ebeling, Barrett 
\& Donovan 2004; Kodama et al.\ 2001).
Beyond a few virial radii from galaxy clusters centres, FOGs
are predicted to weave a complex, web- or sponge-like tapestry
that gives surveys like SDSS, 6dFGS, 2dFGRS \& LCRS their distinctive
appearance (Figure~\ref{fig:twodf}; see also the 2dFGRS homepage
at www.mso.anu.edu.au/2dFGRS ).  

%

We also know that FOGs are highly important
for the mass budget of the Universe (e.g.\ Colberg et al.\ 1999).
Indeed, Cen \& Ostriker (1999) show that for a $\Lambda$ cold dark
matter ($\Lambda$CDM) Universe, a large fraction, perhaps as
much as half (Fukugita, Hogan \& Peebles 1998), 
of baryonic material will not have been observed
as it is situated in the inter-cluster media in a hot and
tenuous gaseous phase.  Along with the dark matter component
and perhaps up to a quarter of the galaxian population, these
baryons are preferentially situated in (inter-cluster) FOGs.  
Moreover, FOGs can provide tests of structure formation
(c.f.\ Colberg, Krughoff \& Connolly 2004 with Pimbblet, Drinkwater
\& Hawkrigg 2004) and cluster evolution (see Colberg et al.\ 1999).
They can also be useful in ascertaining the homogeneity
scale of our Universe (if, indeed one considers there to
be a homogeneity scale; e.g.\ Coleman \& Pietronero 1992
and references therein).  Certainly, given that objects with
scale lengths $>150$ $h^{-1}_{100}$ Mpc exist
and are not chance superpositions, we should
be questioning the validity of the cosmological assumption
up to such lengths and beyond.

The rest of this paper plans out as follows.  In Section~2, we
investigate the numerous methods that one can employ to detect
FOGs and explore their relative merits.  In Section~3 we discuss
the findings and present our conclusions.

\section{Detection}

Already we have seen a number of properties of FOGs.  If sufficiently
large, then they can cause a decrement in the cosmic microwave
background radiation.  They also possess multi-wavelength visibility
(visual; X-ray; etc.).  We describe below how one may take advantage of
such properties to detect them in a given dataset.

\subsection{Optical Overdensity}
\begin{figure}
\vspace*{-0.6in}  
\centerline{\psfig{file=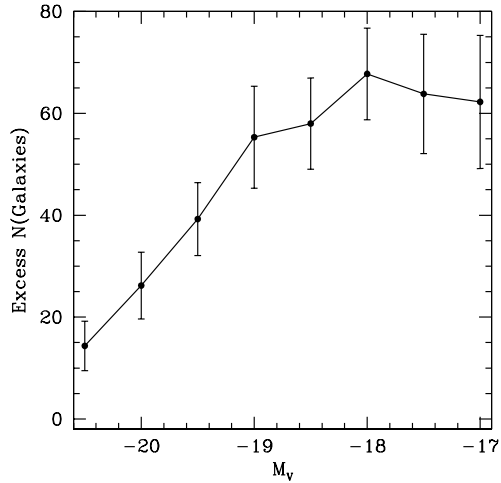,angle=0,width=3.5in}}
\vspace*{-0.4in}  
\caption{\small{Number of galaxies in excess of the field
population (i.e.\ the filament population) as a function of
absolute magnitude (adapted from the investigation
of Pimbblet \& Drinkwater 2004).  
The $\pm 1 \sigma$ errorbars come from a consideration
of Poissonian errors and the variance galaxy number density.
Whilst there are only a few brighter galaxies, the FOG 
contains many more fainter ones.
}}
  \label{fig:overdensity}
\end{figure}

At a very simple level, a FOG is merely an overdensity of galaxies 
compared to the local field\footnote{Here,
we use the term `field' to mean the `average background' level.}
(or void) level.  Pimbblet \& Drinkwater
(2004) use this fact to find a relatively short ($\sim 6 h^{-1}_{100}$ Mpc)
filament between the two close (both in redshift and spatially)
galaxy clusters ACO1079 and ACO1084.  
Mathematically, one can readily compute this galaxy excess as
\begin{equation}
N_{filament} = N_{filament + field} - N_{field}
\end{equation}
Should the observed field sample be too close to the observed
filament sample it will obviously contain some (small but
non-negligible) amount of contamination:
\begin{equation}
N_{field}^{'} = N_{field} + \gamma N_{filament}
\end{equation}
where $\gamma$ is the ratio between the galaxy densities of the
filament and field populations (Paolillo et al.\ 2001).
Substituting $N_{field}^{'}$ instead of $N_{field}$ from
Eq.\ (2) into Eq.\ (1) gives
\begin{equation}
N_{filament}^{'} = N_{filament} (1 - \gamma)
\end{equation}
In Figure~\ref{fig:overdensity} we plot an adaptation of 
the result obtained by Pimbblet \& Drinkwater (2004) utilizing
this particular method.
Most of the excess galaxy population is faint, with only a
few brighter members.  Moreover, only a small fraction ($\approx 30$
per cent) of these galaxies have colours consistent with early-type
galaxies from the two clusters colour-magnitude relations
(Pimbblet et al.\ 2002).  
In shallow (perhaps, mono-colour)
surveys with no supporting redshift information, therefore, 
such an approach is probably not very efficient nor 
exceedingly sensitive and may be somewhat prone to large errors.

%
%

\subsection{X-ray}
Given that a non-negligible fraction of baryonic material
in a $\Lambda$CDM Universe may exist as hot inter-cluster gas, 
one can also consider looking for FOGs in X-ray band passes.
Using the {\it ROSAT} All-Sky Survey data, Briel \& Henry (1995) 
attempted just this by combining together the inter-cluster
regions of 40 cluster pairs. Although they failed to find
any X-ray emitting FOG, they did place an upper limit
on the X-ray surface brightness of $4\times10^{-16}$ ergs 
cm$^{\rm{-2}}$ s$^{\rm{-1}}$ (0.5 -- 2.0 keV).

Scharf et al.\ (2000) make a $5\sigma$ 
joint X-ray/optical detection of $>12 h^{-1}_{50}$ Mpc
(0.5 deg) FOG with a surface brightness of
$1.6\times10^{-16}$ ergs cm$^{\rm{-2}}$ s$^{\rm{-1}}$.
The count rate for this filament, however, is $2-3\sigma$
above background levels.

In the Shapley supercluster meanwhile, Kull \& B{\" o}hringer (1999)
find a promising extended X-ray emission between a close cluster pair
that is $\sim 2.5$ times brighter than Briel \& Henry's (1995)
bound. 
The only problem here is that Scharf et al.\ (2000) note that 
the X-ray emission could be ejecta due to the clusters interacting
(merging) with one another rather than material falling in from 
an actual filament.

More promising progress on this front has been made by 
Tittley \& Henriksen (2001) and Durret et al.\ (2003).
The former detect a gaseous FOG between ACO3391 and ACO3395
that has a minimum flux of $1.3 \times 10^{-12}$ ergs cm$^{\rm{-2}}$ 
s$^{\rm{-1}}$ (0.8--10 keV) and represents at least 2 per cent
of the total mass of the system.  The latter team
study the ACO85 cluster complex and find a highly elongated 
filamentary structure.  Again, however, this filament may the
result of cluster interactions and not a true FOG in the
large-scale structure sense.

It would seem that whilst one may expect there to be significant
X-ray emission from baryonic material contained in FOGs, we
are not quite detecting it with sufficient regularity or 
confidence to use X-ray emission as the primary tool for
FOG detection (unlike in the case of galaxy clusters where X-ray
detections are made with much more confidence; e.g.\ 
Ebeling et al.\ 1996).

\subsection{Lensing}
Pogosyan et al.\ (1998) point out that FOGs that connect
together neighbouring galaxy clusters should have sufficient 
surface mass density as to be detectable in the weak lensing regime.
Indeed, weak lensing would only depend upon the projected density 
and not the square of the projected density like X-rays are
(Pogosyan et al.\ 1998) and therefore it may 
constitute an altogether better way of detecting and
defining FOGs.  There are a small number of investigations
that have been proceeding in this direction.

Kaiser et al.\ (1998) perform a lensing analysis on the
$z=0.4$ supercluster MS 0302+17 and find FOG between two 
of its three component galaxy clusters.  The detection has
remained dubious, however, as there may be foreground structure
interfering, perhaps some edge of chip effects
and residual systematics in the point spread function  
anisotropy correction involved (Gavazzi et al.\ 2004).
Indeed, Gavazzi et al.\ (2004) report that they 
cannot independently confirm
the detection of this particular FOG.

Meanwhile, in other investigations,
Clowe et al.\ (1998) report on the detection of a FOG extending
from the $z=0.8$ rich cluster RXJ 1716+67.  However, the 
size of their imaging is small and it is thus unknown how
far this filament extends in the direction of a nearby cluster.
Gray et al.\ (2002) examines the ACO901/902 supercluster and
find a FOG present.  The significance of the detection is, however,
small.  Superposed with this is the issue that the filament lies in 
the inter-chip region of the analysis.  Nonetheless, this remains
a relatively good detection when compared to the problems 
that Kaiser et al.\ (1998) encounter.

Yet to date, arguably
one of the best weak lensing FOG detections has to 
be that of Dietrich et al.\ (2004) between ACO222 and ACO223.
Not only is the filament detected by weak lensing, but Dietrich
et al.\ (2004) supply supporting evidence from X-ray emission
and increased galaxy density between the clusters.  
Dietrich et al.\ (2004) point out, however, that they could
not find an objective way to define their filament and in this
respect their filament candidate is not very different to that 
of Kaiser et al.\ (1998).

So, it would appear that given imaging of sufficient quality and depth,
weak lensing could provide an excellent way of detecting FOGs,
most especially in combination with other methods (e.g.\ X-ray; 
see above).

\subsection{Redshifts}

\begin{figure}
\vspace*{-0.6in}  
\centerline{\psfig{file=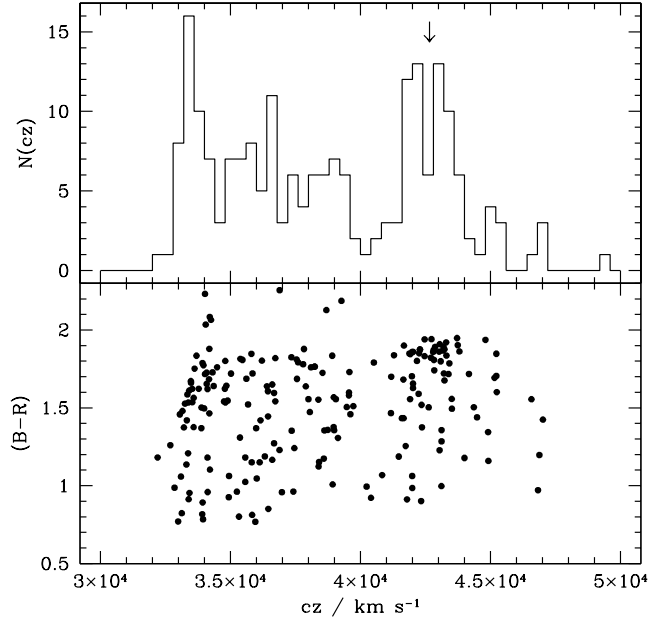,angle=0,width=3.5in}}
\caption{\small{The wall in front of ACO22 discovered by Pimbblet,
Edge \& Couch (2005).  The upper panel shows the redshift distribution
of the cluster (right-hand peak; downward arrow) and the wall 
(left-hand side; $cz<40000$ km s$^{-1}$).  The lower panel shows how
the $(B-R)$ colours of the galaxies vary with redshift.  The colour
distribution and Butcher-Oemler blue fraction is statistically the
same (within $2\sigma$) between the cluster and the wall 
when appropriate magnitudinal cuts are made (Pimbblet, 
Edge \& Couch 2005).
}}
  \label{fig:pec}
\end{figure}

Redshifts of regions around galaxy clusters can provide 
concrete determinations of the presence of FOGs.  
For example, Ebeling, Barrett \& Donovan (2004) report on
a 4 $h^{-1}_{70}$ Mpc filament that is feeding the growth
of the massive cluster MACS J0717.5+3745 at $z=0.55$.  
Its extent beyond
the virial radius of the cluster means that it cannot be the
remnant of some previous interaction or merger whilst its colours
are quite consistent with the colour-magnitude relation
(CMR; e.g.\ Visvanathan \& Sandage 1977; Bower, Lucey \& Ellis 1992)
Indeed, the CMR and other photometric redshift techniques can
also help to better define FOGs.  Kodama et al.\ (2001)
report several `octopus'-like tentacles around ACO851
($z=0.41$) which have colours entirely
consistent with the CMR of the cluster itself.
Pimbblet, Edge \& Couch (2005) locate a large scale wall
covering at least 40 $h^{-1}_{100}$ Mpc situated in
front of ACO22 ($z=0.14$; Figure~\ref{fig:pec}).  
Not only does this wall exhibit a CMR similar to ACO22, but it 
also has a Butcher \& Oemler (1984; see Pimbblet 2003 for a 
review of the Butcher-Oemler effect) blue fraction that does not
change significantly between the cluster and the wall
(Figure~\ref{fig:pec}).

%
%

The above are examples of FOGs around individual clusters.
Of course, it is the modern
large redshift surveys such as 2dFGRS, 6dFGS, SDSS and LCRS that 
are providing
the community with an unprecedented view of the very large scale 
structure of the Universe (e.g.\ Figure~\ref{fig:twodf}).  
With such large datasets, finding individual 
filaments can become as easy as
looking at the regions between two galaxy clusters in three 
dimensional space and making an appropriate cut at some 
galaxy density threshold\footnote{Pimbblet, Edge \& Couch (2005)
note that the typical surface density of FOGs is of the
order 10 bright (brighter than say $M^{*}+2$) 
galaxies per square h$^{-1}$ Mpc.} 
to determine if there is a significant
overdensity of galaxies present (e.g.\ Pimbblet, Drinkwater
\& Hawkrigg 2004). 
One potential pitfall is that one may mistake
a redshift space distortion (see Hawkins et al.\ 2003)
for a FOG.  Pimbblet, Drinkwater \& Hawkrigg (2004) 
circumvent this by only considering cluster-cluster
pairs within 1000 kms$^{-1}$ of each other and check
the FOG distribution angles along the line of sight
to ensure that no `fingers of god' are mistaken for
FOGs (and conversely, no end-on FOG is mistaken for a
cluster!).
Also, surveys like 2dFGRS only cream off the very luminous galaxies
(and even at bright magnitudes, are incomplete; Cross et al.\ 2004).  
They tell us little about the low surface
brightness populations which likely contribute a non-negligible
fraction of any FOG's mass.  The next-generation of deeper redshift
surveys (e.g.\ using AA$\Omega$ on the Anglo-Australian Telescope)
should help us to address this point.  

We note in passing, however, that one potentially unanswered 
question is whether a FOG
can exist that is not connected to any cluster?
If they can, this would bias the results of these kind
of investigations which focus exclusively on intercluster regions
as the locations to search out FOGs.
Recent work by Fairall et al.\ (2004) suggests, however, 
that there can be no isolated FOGs.

\subsection{Further Statistical Methods}
Given that the large redshift surveys may contain several hundred
FOGs, it can be better to approach the entire FOG population 
in a more consistent (and less time-consuming) manner.
There are a multitude of statistics available for the 
analysis of large-scale structure and FOGs, and here we only
review a small, representative selection of these approaches.

\begin{figure}
\vspace*{-0.58in}  
\centerline{\psfig{file=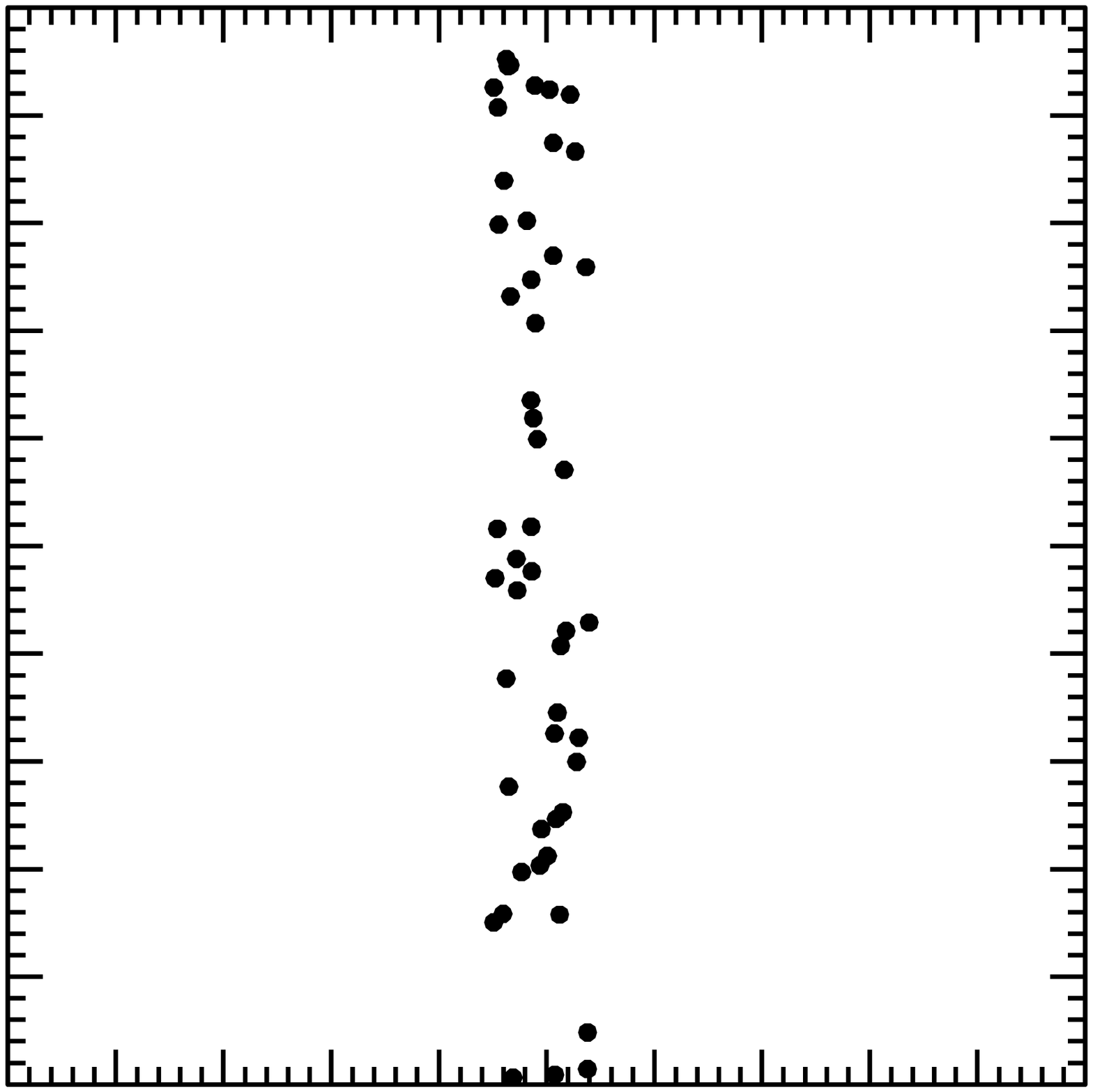,angle=0,width=2.37in}
\psfig{file=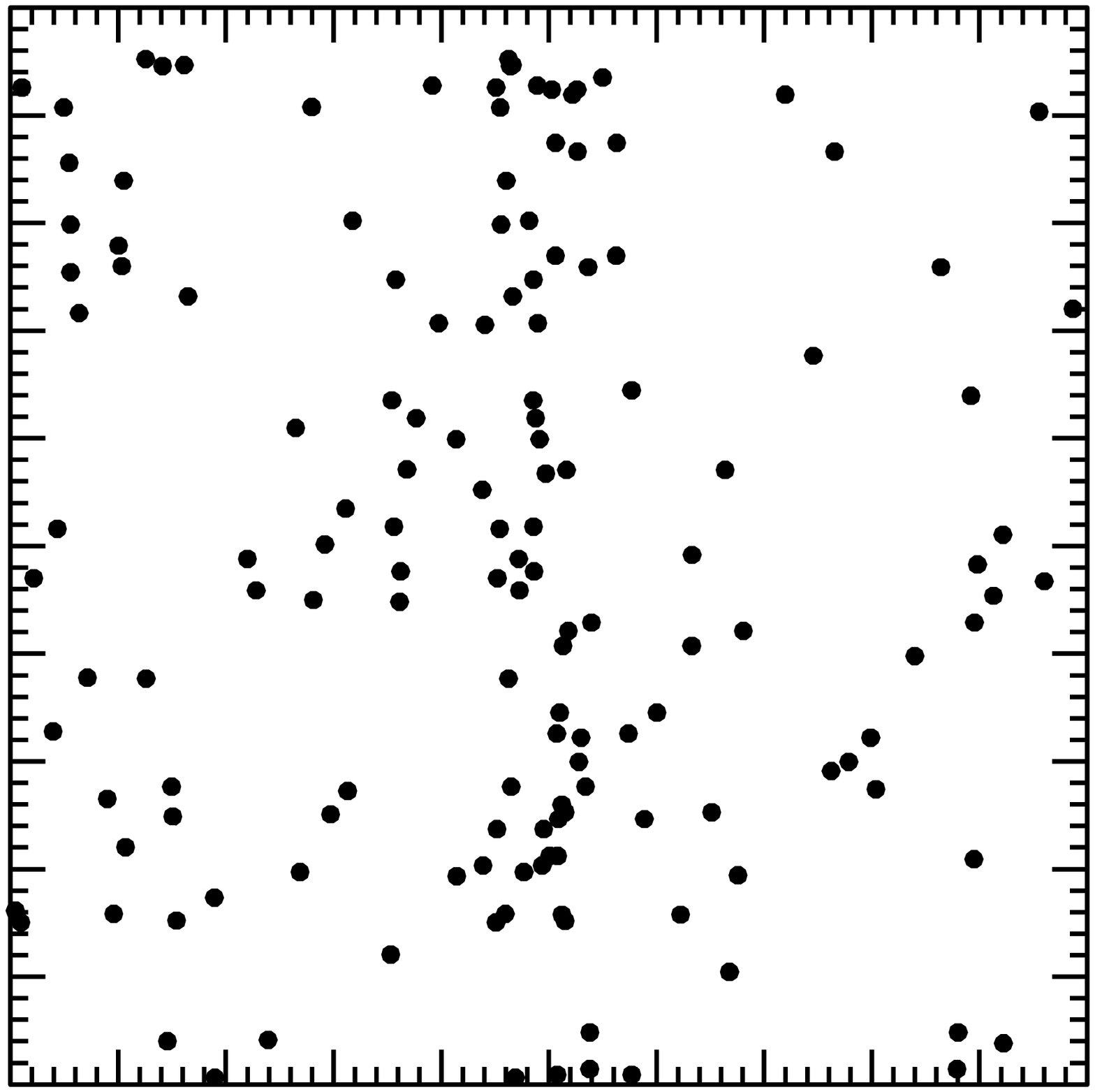,angle=0,width=2.37in}}
\hspace*{0.03in}
\centerline{\psfig{file=smooth1.ps,angle=0,width=2in}
\hspace*{0.32in}
\psfig{file=smooth2.ps,angle=0,width=2in}}
\caption{\small{Top Left: a simulated toy-model
FOG as might be found in a redshift survey;
generated as randomly sampled points along a vertical (thick) line.
Bottom Left: a smoothed
distribution of this FOG.  An isodensity contour cut could
readily be employed to delineate it. 
Top Right: adding in random noise 
results in the original signal becoming harder to pick out -- a 
variant of the so-called connect the dots type of problem.
Bottom Right: without the noise, the isodensity contours are approximately
straight, but with noise, they start to look much more curvy
Further, it may also appear that at lower densities there is a
horizontal FOG passing through the overdensity in the lower 
portion of the plot;
but it should still be possible to detect the original FOG by 
thresholding at a particular isodensity contour.
}}
  \label{fig:ctd}
\end{figure}

Some readers may be familiar with the children's game `connect
the dots' (CTD herein).  The idea is to connect a series of points
on a plane, in a particular sequence,
in order to form some kind of picture at the end of it.
One well known relative of CTD is the travelling salesman
problem where the challenge is to connect some points in
by shortest possible route.  Yet
finding FOGs in redshift surveys like 2dFGRS can also be thought
of as another variant of CTD types of games (Arias-Castro et al.\ 2004).  
Consider a toy-model: a two dimensional
curve that one randomly\footnote{Random numbers are generated 
using the method of Pimbblet \& Bulmer (2005).}
samples points from along its length 
(i.e.\ galaxies contained in a FOG).  Add in some random 
noise, and now the problem becomes whether we are able to recover 
the original curve (FOG) in the presence of clutter (field galaxies;
chance superpositions; etc.).  
Donoho and collaborators (e.g.\ Arias-Castro et al.\ 2004) 
approach the CTD problem from 
a number of angles.  
One promising method is to 
make use of a multiscale adaptive geometric analysis.
By using a zoo of parallelogram strips of various angles,
lengths, widths and eccentricities one can evaluate (count) 
the galaxy population in all such strips.  Then all one
has to do is identify strips with unusually high count 
rates and search for long runs of such strips that would
constitute a good continuation of a given curve.  
The problem with this analogy is that 
FOGs are not perfect lines or curves: they can be bumpy, lumpy and clumpy 
in three dimensions (Figure~\ref{fig:ctd}).  Finding a FOG that is only
just above the density of the clutter can prove to be hard
(Figure~\ref{fig:ctd}), especially if one is interested in the
morphology of FOGs.

%
%

\begin{figure}
\vspace*{-0.6in}  
\centerline{\psfig{file=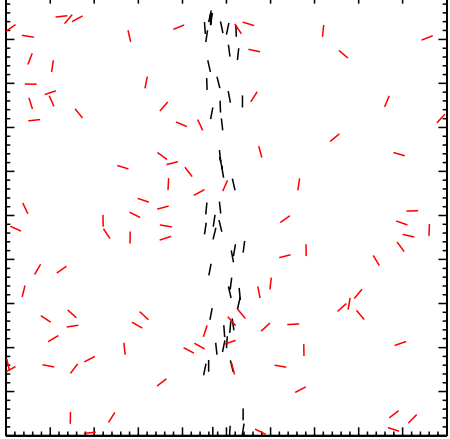,angle=0,width=2.75in}}
\vspace*{-0.3in}  
\caption{\small{If one makes use of galaxy angles (i.e.\ vectors), then
the problem posed by Figure~\ref{fig:ctd} potentially becomes much
easier to solve.  Here, the members of the FOG have orientations
of $\theta=90 \pm 15$ deg, whereas the interlopers (red) 
have purely random orientation angles.
}}
  \label{fig:ang}
\end{figure}

However, if one can make use of galaxy orientation angles
(Pimbblet 2005), this problem now becomes vectorized 
(a so-called `connect-the-darts' problem)
and potentially easier to solve (Figure~\ref{fig:ang}; 
Arias-Castro et al.\ 2004).
We know from early work by Binggeli (1982)
that the major axis of 
galaxy clusters are generally aligned exceptionally 
well with their first-ranked (usually a cD-type) galaxy 
and that close cluster pairs generally point 
to each other.  Moreover in $\Lambda$CDM,
filamentary structure funnels material
along preferred directions toward clusters.  
Since galaxy alignment tends to follow the orientation of
clusters and the filaments that feed them, 
Pimbblet (2005; and references therein) took 
advantage of this fact by
computing the degree of galaxy angle anisotropy for a selected
region of the sky.  Regions that have significantly anisotropic
distributions of galaxy angles (as in Figure~\ref{fig:ang})
readily show up.

%
%

Since it is easy to think of galaxies as points within large-scale
surveys, it is no surprise that one can apply many mathematical 
approaches to delineating FOGs and other intrinsic patterns within them.  
A popular approach is to use a minimal spanning tree formalism (MST 
herein; e.g.\ Barrow, Bhavsar \& Sonoda 1985; Bhavsar \& 
Ling 1988; Krzewina \& Saslaw 1996; Doroshkevich et al.\ 2000;
Doroshkevich et al.\ 2004 amongst others).  One draw back of MST
is that it will produce a unique graph for a given set of points.
It is known that large-scale surveys such as SDSS and
2dFGRS are incomplete by about 10 to 20 per cent at all magnitudes
(Cross et al.\ 2004; Pimbblet et al.\ 2001) -- a fact that makes MST
potentially a poorer choice for analyzing large scale structure 
with than other methods.

Such other statistical methods, that here we will mention 
only briefly, include the use of Voronoi 
(and the complementary Delaunay) tessellations (e.g.\ van de Weygaert 1994).
Essentially, the Voronoi tessellation can be thought of as
constructing a skeleton of the Universe by
simply finding the bisection line between a single point and 
every other point.  This process is
repeated for each point and then the Voronoi tessellation is
then the unification of all the halfplanes that have 
been created (see van de Weygaert 1994 for more indepth detail).
Analysis of the cell-like structures of the Voronoi skeleton
can inform one about the underlying galaxy distribution, although direct
detection of FOGs from these tessellations remains a relatively
unattempted task.
Minkowski functionals such as {\sc shapefinders} 
(e.g.\ Bharadwaj et al.\ 2000; Pandey \& Bharadwaj 2005;
see also Shandarin, Sheth \& Sahni 2004) and
the genus statistics (e.g.\ Hoyle et al.\ 2002; Hoyle, Vogeley \& 
Gott 2002) can also provide us with a direct way of 
analyzing the structure of the galaxy distribution.
Moreover, they can also provide a direct measure of the 
`filamentarity' and `planarity' of the Universe 
(e.g.\ Schmalzing et al.\ 1999) and one can readily 
delineate FOGs from them by using an isodensity contour cut.
Finally, we should also mention that there are a 
host of other marked point processes (Stoica et al.\ 2005 and
references therein) 
which are also capable of recovering individual FOGs.
All of these methods highlight the presence of FOGs within
redshift surveys to varying degrees. 

\section{Discussion and Conclusions}

Given the above methods to detect FOGs, there is a large
amount of literature dedicated to their dissemination.
One issue that seems to be prominent in the literature is
the mixed nomenclature for FOGs.  Many authors refer to
them as `walls' (Geller \& Huchra 1989), others call them
`filaments', some use the term `sheets' or `pancakes'.
So what is the difference between all these terms?
Pimbblet, Drinkwater \& Hawkrigg (2004) and Colberg,
Krughoff \& Connolly (2004) attempt to refine these
definitions by dividing detected FOGs into several categories
based upon their visual morphology.  So the difference
between a filament and a wall then becomes a matter of
how thick (or, equally, how wide) the FOG is in 
three-dimensional space
(i.e.\ a filament will have depth $\approx$ width).
Sheets are then synonymous with walls.
Is this kind of morphological classification useful?  
Given that walls appear to be much, much rarer than `normal'
filaments (Pimbblet, Drinkwater \& Hawkrigg 2004) 
and unlike filaments, they do not possess non-isotropic galaxy
orientations (Pimbblet 2005) -- yes.  Their relative abundances
(also filling factors) and lengths
should help us to better constrain the ideal cosmological 
paradigm (e.g.\ in $\Lambda$ cold dark matter
cosmologies, studies of FOGs can readily exclude bias parameters
of $b>1.5$; Bharadwaj \& Pandey 2004) as can the number of FOGs 
connected to clusters of a given mass (Colberg, Krughoff \& Connolly
2004).

At the outset of this work, the question `what is a filament of
galaxies?' was posed.  
This work has reviewed a number of methods for finding,
detecting and defining FOGs in datasets of varying complexity.  
Those that are more physically motivated (gravitational weak
lensing searches; X-ray searches) appear to be an optimal
way to detecting them (especially in unison), but yet, 
they remain a time-intensive method owing to the required
amount of observing time to get down to sufficient limiting
magnitudes and fluxes.  

We have also investigated how FOGs are detected in large redshift
surveys using a variety of methods ranging from simple isodensity
thresholding to more involved statistics like the MST.  
Here, it seems that even the very simple approaches can yield
useful results, such as the distribution and abundances of 
FOG lengths, that are in remarkable agreement with theory.

\section*{Acknowledgments} 

KAP thanks Michael Drinkwater, Alastair Edge and Mary Hawkrigg for 
their support.  This work is financed through the award of a
University of Queensland EPSA Research Fellowship and
a UQRSF grant.

On a final note, I also wish to express my sincere gratitude to 
the two referees, Dominique Proust and Tony Fairall, who provided 
quick and useful reports that have improved the quality of this work.



\begin{thebibliography}{}

\bibitem[\protect\citeauthoryear{Abazajian et 
al.}{2004}]{2004AJ....128..502A} Abazajian K., et al., 2004, AJ, 128, 502

\bibitem[\protect\citeauthoryear{Abell}{1965}]{1965ARA&A...3....1A} Abell 
G.~O., 1965, ARA\&A, 3, 1 

\bibitem[\protect\citeauthoryear{Arias-Castro et al.}{2004}]{2004...AC}
Arias-Castro E., Donoho D., Huo X., Tovey C., 2004,
Advances in Applied Probability, submitted
(www-stat.stanford.edu / $\sim$donoho / Reports / 2004 / CTD-Arias-etal.pdf)

\bibitem[\protect\citeauthoryear{Atrio-Barandela \& 
Kashlinsky}{1992}]{1992ApJ...390..322A} Atrio-Barandela F., Kashlinsky A., 
1992, ApJ, 390, 322 

\bibitem[\protect\citeauthoryear{Barrow, Bhavsar, \& 
Sonoda}{1985}]{1985MNRAS.216...17B} Barrow J.~D., Bhavsar S.~P., Sonoda 
D.~H., 1985, MNRAS, 216, 17 

\bibitem[\protect\citeauthoryear{Bharadwaj \& 
Pandey}{2004}]{2004ApJ...615....1B} Bharadwaj S., Pandey B., 2004, ApJ, 
615, 1 

\bibitem[\protect\citeauthoryear{Bharadwaj et 
al.}{2000}]{2000ApJ...528...21B} Bharadwaj S., Sahni V., Sathyaprakash 
B.~S., Shandarin S.~F., Yess C., 2000, ApJ, 528, 21 

\bibitem[\protect\citeauthoryear{Bhavsar \& 
Ling}{1988}]{1988ApJ...331L..63B} Bhavsar S.~P., Ling E.~N., 1988, ApJ, 
331, L63 

\bibitem[\protect\citeauthoryear{Binggeli}{1982}]{1982A&A...107..338B} 
Binggeli B., 1982, A\&A, 107, 338 

\bibitem[\protect\citeauthoryear{Bond et al.}{1996}]{1996Nature} 
Bond J.~R., Kofman L., Pogosyan, D., 1996, Nature, 380, 603

\bibitem[\protect\citeauthoryear{Bower, Lucey, \& 
Ellis}{1992}]{1992MNRAS.254..601B} Bower R.~G., Lucey J.~R., Ellis R.~S., 
1992, MNRAS, 254, 601 

\bibitem[\protect\citeauthoryear{Briel \& 
Henry}{1995}]{1995A&A...302L...9B} Briel U.~G., Henry J.~P., 1995, A\&A, 
302, L9 

\bibitem[Butcher \& Oemler (1984)]{1984ApJ...285..426B} Butcher, H.\ \& 
Oemler, A.\ 1984, ApJ, 285, 426 

\bibitem[\protect\citeauthoryear{Cen \& 
Ostriker}{1999}]{1999ApJ...514....1C} Cen R., Ostriker J.~P., 1999, ApJ, 
514, 1 

\bibitem[\protect\citeauthoryear{Chodorowski}{1994}]{1994MNRAS.266..897C} 
Chodorowski M., 1994, MNRAS, 266, 897 

\bibitem[\protect\citeauthoryear{Clowe et al.}{1998}]{1998ApJ...497L..61C} 
Clowe D., Luppino G.~A., Kaiser N., Henry J.~P., Gioia I.~M., 1998, ApJ, 
497, L61 

\bibitem[\protect\citeauthoryear{Colberg et 
al.}{2000}]{2004MNRAS.C} Colberg J.~M., Krughoff K.~S., Connolly A.~J., 
2004, MNRAS in press (astro-ph/0406665)

\bibitem[\protect\citeauthoryear{Colberg et 
al.}{2000}]{2000MNRAS.319..209C} Colberg J.~M., et al., 2000, MNRAS, 319, 
209 

\bibitem[\protect\citeauthoryear{Colberg et 
al.}{1999}]{1999MNRAS.308..593C} Colberg J.~M., White S.~D.~M., Jenkins A., 
Pearce F.~R., 1999, MNRAS, 308, 593 

\bibitem[\protect\citeauthoryear{Coleman \& 
Pietronero}{1992}]{1992PhR...213..311C} Coleman P.~H., Pietronero L., 1992, 
PhR, 213, 311 

\bibitem[\protect\citeauthoryear{Cross et al.}{2004}]{2004MNRAS.349..576C} 
Cross N.~J.~G., Driver S.~P., Liske J., Lemon D.~J., Peacock J.~A., Cole 
S., Norberg P., Sutherland W.~J., 2004, MNRAS, 349, 576 

\bibitem[\protect\citeauthoryear{Dietrich et 
al.}{2004}]{2004astro.ph..6541D} Dietrich J.~P., Schneider P., Clowe D., 
Romano-Diaz E., Kerp J., 2004, preprint, astro-ph/0406541 

\bibitem[\protect\citeauthoryear{Doroshkevich et 
al.}{2000}]{2000MNRAS.315..767D} Doroshkevich A.~G., Fong R., McCracken 
H.~J., Ratcliffe A., Shanks T., Turchaninov V.~I., 2000, MNRAS, 315, 767 

\bibitem[\protect\citeauthoryear{Doroshkevich et 
al.}{2004}]{2004A&A...418....7D} Doroshkevich A., Tucker D.~L., Allam S., 
Way M.~J., 2004, A\&A, 418, 7 

\bibitem[\protect\citeauthoryear{Durret et al.}{2003}]{2003A&A...403L..29D} 
Durret F., Lima Neto G.~B., Forman W., Churazov E., 2003, A\&A, 403, L29 

\bibitem[\protect\citeauthoryear{Ebeling, Barrett, \& 
Donovan}{2004}]{2004ApJ...609L..49E} Ebeling H., Barrett E., Donovan D., 
2004, ApJ, 609, L49 

\bibitem[\protect\citeauthoryear{Ebeling et 
al.}{1996}]{1996MNRAS.281..799E} Ebeling H., Voges W., Bohringer H., Edge 
A.~C., Huchra J.~P., Briel U.~G., 1996, MNRAS, 281, 799 

\bibitem[\protect\citeauthoryear{Fairall et 
al.}{2004}]{2004astro.ph.11437F} Fairall A., Turner D., Pretorius M.~L., 
Wiehahn M., McBride V., de Vaux G., Woudt P.~A., 2004, preprint, 
astro-ph/0411437 

\bibitem[\protect\citeauthoryear{Fukugita, Hogan, \& 
Peebles}{1998}]{1998ApJ...503..518F} Fukugita M., Hogan C.~J., Peebles 
P.~J.~E., 1998, ApJ, 503, 518 

\bibitem[\protect\citeauthoryear{Gavazzi et 
al.}{2004}]{2004A&A...422..407G} Gavazzi R., Mellier Y., Fort B., 
Cuillandre J.-C., Dantel-Fort M., 2004, A\&A, 422, 407 

\bibitem[\protect\citeauthoryear{Geller \& 
Huchra}{1989}]{1989Sci...246..897G} Geller M.~J., Huchra J.~P., 1989, Sci, 
246, 897 

\bibitem[\protect\citeauthoryear{Gray et al.}{2002}]{2002ApJ...568..141G} 
Gray M.~E., Taylor A.~N., Meisenheimer K., Dye S., Wolf C., Thommes E., 
2002, ApJ, 568, 141 

\bibitem[\protect\citeauthoryear{Hawkins et 
al.}{2003}]{2003MNRAS.346...78H} Hawkins E., et al., 2003, MNRAS, 346, 78 

\bibitem[\protect\citeauthoryear{Hoyle et al.}{2002}]{2002ApJ...580..663H} 
Hoyle F., et al., 2002, ApJ, 580, 663 

\bibitem[\protect\citeauthoryear{Hoyle, Vogeley, \& 
Gott}{2002}]{2002ApJ...570...44H} Hoyle F., Vogeley M.~S., Gott J.~R.~I., 
2002, ApJ, 570, 44 

\bibitem[Huchra et al.(1983)]{1983ApJS...52...89H} Huchra, J., Davis, M., 
Latham, D., \& Tonry, J.\ 1983, ApJS, 52, 89 

\bibitem[\protect\citeauthoryear{Jenkins et 
al.}{1998}]{1998ApJ...499...20J} Jenkins A., et al., 1998, ApJ, 499, 20 

\bibitem[\protect\citeauthoryear{Jones et al.}{2004}]{2004MNRAS.355..747J} 
Jones D.~H., et al., 2004, MNRAS, 355, 747 

\bibitem[\protect\citeauthoryear{Kaiser et al.}{1998}]{1998..K}
Kaiser N., Wilson G., Luppino G., Kofman L., Gioia I., Metzger M., 
Dahle H., 1998, preprint, astro-ph/9809268 

\bibitem[\protect\citeauthoryear{Katz et al.}{1996}]{1996ApJ...457L..57K} 
Katz N., Weinberg D.~H., Hernquist L., Miralda-Escude J., 1996, ApJ, 457, 
L57 

\bibitem[\protect\citeauthoryear{Kodama et al.}{2001}]{2001ApJ...562L...9K} 
Kodama T., Smail I., Nakata F., Okamura S., Bower R.~G., 2001, ApJ, 562, L9 

\bibitem[\protect\citeauthoryear{Krzewina \& 
Saslaw}{1996}]{1996MNRAS.278..869K} Krzewina L.~G., Saslaw W.~C., 1996, 
MNRAS, 278, 869 

\bibitem[\protect\citeauthoryear{Kull \& B{\" 
o}hringer}{1999}]{1999A&A...341...23K} Kull A., B{\" o}hringer H., 1999, 
A\&A, 341, 23 

\bibitem[\protect\citeauthoryear{Pandey \& 
Bharadwaj}{2005}]{2005MNRAS.357.1068P} Pandey B., Bharadwaj S., 2005, 
MNRAS, 357, 1068 

\bibitem[\protect\citeauthoryear{Paolillo et 
al.}{2001}]{2001A&A...367...59P} Paolillo M., Andreon S., Longo G., Puddu 
E., Gal R.~R., Scaramella R., Djorgovski S.~G., de Carvalho R., 2001, A\&A, 
367, 59 

\bibitem[\protect\citeauthoryear{Pimbblet}{2005}]{2005MNRAS.358..256P} 
Pimbblet K.~A., 2005, MNRAS, 358, 256

\bibitem[\protect\citeauthoryear{Pimbblet, Edge, \& 
Couch}{2005}]{2005MNRAS.357L..45P} Pimbblet K.~A., Edge A.~C., Couch W.~J., 
2005, MNRAS, 357, L45 

\bibitem[\protect\citeauthoryear{Pimbblet \& 
Bulmer}{2005}]{2005PASA...22....1P} Pimbblet K.~A., Bulmer M., 2005, PASA, 
22, 1 

\bibitem[\protect\citeauthoryear{Pimbblet, Drinkwater, \& 
Hawkrigg}{2004}]{2004MNRAS.354L..61P} Pimbblet K.~A., Drinkwater M.~J., 
Hawkrigg M.~C., 2004, MNRAS, 354, L61 

\bibitem[\protect\citeauthoryear{Pimbblet \& 
Drinkwater}{2004}]{2004MNRAS.347..137P} 
Pimbblet K.~A., Drinkwater M.~J., 2004, MNRAS, 347, 137 

\bibitem[\protect\citeauthoryear{Pimbblet}{2003}]{2003PASA...20..294P} 
Pimbblet K.~A., 2003, PASA, 20, 294 

\bibitem[\protect\citeauthoryear{Pimbblet et 
al.}{2002}]{2002MNRAS.331..333P} Pimbblet K.~A., Smail I., Kodama T., Couch 
W.~J., Edge A.~C., Zabludoff A.~I., O'Hely E., 2002, MNRAS, 331, 333 

\bibitem[\protect\citeauthoryear{Pimbblet et 
al.}{2001}]{2001MNRAS.327..588P} Pimbblet K.~A., Smail I., Edge A.~C., 
Couch W.~J., O'Hely E., Zabludoff A.~I., 2001, MNRAS, 327, 588 

\bibitem[\protect\citeauthoryear{Pogosyan et 
al.}{1998}]{1998wfsc.conf...61P} Pogosyan D., Bond J.~R., Kofman L., 
Wadsley J., 1998, in Wide Field Surveys in Cosmology, 
14th IAP meeting held May 26-30, 1998 (Paris: Editions Frontieres), 61 

\bibitem[\protect\citeauthoryear{Scharf et al.}{2000}]{2000ApJ...528L..73S} 
Scharf C., Donahue M., Voit G.~M., Rosati P., Postman M., 2000, ApJ, 528, 
L73 

\bibitem[\protect\citeauthoryear{Schmalzing et 
al.}{1999}]{1999ApJ...526..568S} Schmalzing J., Buchert T., Melott A.~L., 
Sahni V., Sathyaprakash B.~S., Shandarin S.~F., 1999, ApJ, 526, 568 

\bibitem[\protect\citeauthoryear{Shandarin, Sheth, \& 
Sahni}{2004}]{2004MNRAS.353..162S} Shandarin S.~F., Sheth J.~V., Sahni V., 
2004, MNRAS, 353, 162 

\bibitem[\protect\citeauthoryear{Shectman et 
al.}{1996}]{1996ApJ...470..172S} Shectman S.~A., Landy S.~D., Oemler A., 
Tucker D.~L., Lin H., Kirshner R.~P., Schechter P.~L., 1996, ApJ, 470, 172 

\bibitem[\protect\citeauthoryear{Stoica et al.}{2004}]{2004astro.ph..5370S} 
Stoica R.~S., Martinez V.~J., Mateu J., Saar E., 2005, A\&A, in press 
(astro-ph/0405370)

\bibitem[\protect\citeauthoryear{Tittley \& 
Henriksen}{2001}]{2001ApJ...563..673T} Tittley E.~R., Henriksen M., 2001, 
ApJ, 563, 673 

\bibitem[\protect\citeauthoryear{van de 
Weygaert}{1994}]{1994A&A...283..361V} van de Weygaert R., 1994, A\&A, 283, 
361 

\bibitem[\protect\citeauthoryear{Visvanathan \& 
Sandage}{1977}]{1977ApJ...216..214V} Visvanathan N., Sandage A., 1977, ApJ, 
216, 214 

\bibitem[\protect\citeauthoryear{Zeldovich, Einasto, \& 
Shandarin}{1982}]{1982Natur.300..407Z} Zeldovich I.~B., Einasto J., 
Shandarin S.~F., 1982, Nature, 300, 407 

\end{thebibliography}
\end{document}